\documentstyle[12pt]{article}
\begin{document}
\begin{center}
{\Large {\bf KINETICS OF THE WATER ABSORPTION IN SLAG CEMENT MORTARS.}}\\

\vspace{1cm}
Ernesto Villar Coci\~na$^1$, Eduardo Valencia Morales$^1$, Jes\'us Hern\'andez Ruiz$^1$,
Jorge Vega Leyva$^1$ and Julio Antiquera-Mu\~noz$^2$\\

1. Physics Department, Central University of Las Villas, CUBA.\\
2. University of Magallanes (UMAG), Chile.\\
\end{center}

\vspace{1cm}
\begin{abstract}

  We study the kinetics of absorption of water in portland cement mortars added with 60, 70 and 80\% of granulated blast furnace slag (GGBS or SLAG) cured in water and at open air and preheating at 50 and $100^0$C. A mathematical model is presented that allows describing the process not only in early ages where the capillary sorption is predominant but also for later and large times where the diffusive processes through the finer and gel pores are considered. The fitting of the model by computerized methods enables us to determine the parameters that characterize the process: i. e.  the sorptivity coefficient (S) and diffusion coefficient (D). This allows the description of the process for all times and offers the possibility to know the contributions of both, the diffusive and capillary processes. The results show the influence of the curing regime and the preheating temperature on the behavior of GGBS mortars.  \\
\end{abstract}.\\\\\\\\


\newpage
\section{Introduction}

  The role of water in the mechanism of deterioration of porous building materials has been recognized for a long time. The conjugate action of mechanical and chemical damages arising for instance from freeze/thaw cycles, shrinkage-induced cracking, salt crystallization and leaching \cite{Addleson}may lead to rapid degradation of the bulk mechanical properties of building materials and significantly reduce the service life of constructions.   \\
  The durability of building structures is thus critically determined by the rate at which water infiltrate and move through the porous structure. In addition the penetration of water in building materials provides a mechanism and path for the penetration of deleterious materials like chloride and sulfate ions which can lead to the corrosion of steel reinforcement in concretes \cite{Christensen, Jensen, Thoft, Luping, Mironova}. Sorptivity and diffusion are the primary mechanisms that control the transport of deleterious materials inside of the concrete \cite{Martys}. The diffusion corresponds to the ionic movement under the action of a concentration gradient (it occurs fundamentally when the material is dampened) and sorptivity is the property that characterizes the disposition of a porous material of absorbing and to transmit water by capillarity. \\
  Because of its economic implications, the problem of water movements in porous building materials has received great attention in the past and also at present. A better understanding of the water absorption and moisture transfer can therefore reduce or prevent damage in building materials. For that reason these phenomena are of interest to physicist and engineers.\\
  This phenomenon has been the object of intensive research \cite{Martys,1Hall,Ho,2Hall,3Hall,Farrell,Ohdaira,Goual,McCarter1,McCarter2}where different methods of measuring the uptake and movement of water in these materials (gravimetry, microwave method, electrical conductivity, relative humidity, ultrasonic method, nuclear magnetic resonance, etc)have been used.   \\
It is known that the use of finely divided siliceous materials (such as fly ash, ground granulated blast furnace (GGBS or slag), rice husk ash, silica fume, etc) in concrete is the most direct, technically feasible and economically attractive solution to the problem of reinforced concrete durability.\\
In particular, these materials greatly improve the durability of concrete through control of high thermal gradients, pore refinement, depletion of cement alkalis and the capability for continued long-term hydration or pozzolanic reaction \cite{Swamy}. In developing countries where the increasing demand for construction take place, the technical advantages of pozzolans and slag in concrete are complemented by other economic, ecological and environmental considerations. These materials reduce substantially not only the energy consumption in the production of concrete, but also help to reduce environmental pollution through reduced emission of carbon dioxide.\\
Slag is a by-product from the production of steel. During production liquid slag is rapidly quenched from a high temperature by immersion in water \cite{Metha}. The slag is a glassy, granular, non-metallic product that consists essentially of silicates and aluminosilicates of calcium and other bases \cite{Klieger}.\\
The addition of slag (GGBS) diminishes the large pores and reduces porosity. It is due to the hydration of slag \cite{Metha1}which reacts with calcium hydroxide (originated during the hydration of portland cement) to form calcium silicate hydrate (CSH) gel with very small pore size. Thereby the slag increase the strength of the concrete product, which is more marked with the duration of exposure in comparison with the concrete based on ordinary portland cement OPC \cite{Kalid, Mejia, Talero, Thomas}. \\
Ho et. al. \cite{Ho1} showed that concrete quality with response to interrupted curing is improved by the incorporation of slag in concrete. The rate of water absorption by capillary action is significantly decreased when slag is used.  GGBS concrete is substantially more resistant to chloride diffusion than portland cement concrete of the same strength grade. The uses of slag will considerable increase the service life of structures exposed to chloride environment and to provide sufficient cover to the steel reinforcement. Swamy has presented a critical and valuable evaluation about the use of slag in concrete \cite{Swamy}.\\
From the scientific point of view and for numerous applications, simple but accurate models describing the water transport in mortars and concretes are important. Several models (which involve both analytical and numerical models) have been developed for describing this phenomenon in mortars and concretes without and with additions \cite{Martys, 3Hall, Kuntz, Holm, Martys2, Claisse}, but they do not always agree with the experimental results. Several researchers describe the water absorption in concrete in early ages where the process of capillary absorption prevails but they don't describe the complete process where the moisture diffusion through the finer or gel pores may be considered. These finer pores increase its importance with time \cite{Martys}. A model that considers both the capillary and diffusion phenomenon and allows us to explain the process for all ages will be of great importance. \\
The goal of the present paper is to develop a mathematical model for describing the water transport in GGBS cement mortars.  The model proposed here consider that the capillary sorption and diffusion are the principal mechanisms to the absorption of water in slag cement mortars but this don't consider for simplicity an for having a less contribution other mechanism such as permeation, osmosis, thermal migration, etc that can be present. This model allows us to describe the process for early and large times. This gives the possibility of to know the contributions of the diffusion and capillary action in the water transport in cement mortars, which is very useful to predict the penetration of chloride or other ions that deteriorate the steel reinforcement in concretes. The fitting of the model by computerized methods lets us determine process parameters, such as sorptivity coefficient and diffusion coefficient. It allows us to characterize the kinetics of the water transport in GGBS cement mortars with 60, 70 and 80\% of slag. Besides, the influence on the water sorption process of different curing regimes (curing in water and at open air) as well as the effect of changing the preheating temperature is analyzed.\\

\section{EXPERIMENTAL}

  The experimental data presented in this paper were obtained from Mejias and Talero \cite{Mejia, Talero}. \\
The samples used correspond to cylindrical mortars (with a diameter and height of 50.8 and 50 mm respectively) manufactured with OPC (Table 1) blended with GGBS additions of 60, 70 and 80\% by weight. A curing under water and at open air by 15 days was used and a posterior preconditioning of the samples at $100^0$C during 6h and cooling in desiccator for 24 h was applied. Other samples cured under water were dried at $50^0$C during 48 h and cooling in desiccator. Table 2 shows the general data for the samples.\\

\begin{center}
\begin{tabular}{|c|c|c|c|c|}\hline\hline
Cement & $SC_3$ & $SC_2$  & $AC_3$& $FAC_4$\\ \hline\hline

OPC&46.95& 19.98 & 14.69& 5.33\\ \hline

\end{tabular}
\normalsize
\end{center}

{\bf Tabla 1}. Chemical composition of the OPC. \\

One of the ends of the samples was placed on a thin sponge (3 mm) to absorb water. The weight of the specimens was measured intermittently for about two weeks. Figures 1 and 2 show a scheme of the sorption test and the cumulative water absorption versus square root of time respectively. \\
\begin{center}
\begin{tabular}{|c|c|c|c|}\hline\hline
 Sample designations& Compositions  & Curing regime& Preconditioning temperature \\
   &  & (15 days) &     \\   \hline\hline
 S80W100   & OPC + 80\% slag & Water & $100^0$C \\ \hline
 S70W100   & OPC + 70\% slag & Water & $100^0$C \\ \hline
 S60W100   & OPC + 60\% slag & Water & $100^0$C\\ \hline
 S80W50    & OPC + 80\% slag & Water & $50^0$C\\ \hline
 S80A100   & OPC + 80\% slag & Air   & $100^0$C\\ \hline

\end{tabular}
\normalsize
\end{center}
{\bf Table 2}. General data for the samples.\\

\section{Formulation of the problem }

The process of water transport in concrete has been studied by many researchers. In most of the cases has been possible to explain the process satisfactorily for early ages in dry samples and without additions. It is frequently found that if a mortar or concrete surface is exposed to wetting by water then the cumulative water absorption (normalized to the exposed surface area) M/A (in $kg/m^2$) is proportional, during the initial absorption period, to the square root of elapsed wetting time t: \cite{3Hall}\\

\begin{equation}
\frac{M}{A}=S t^\frac{1}{2}
\end{equation}

where S is the sorptivity coefficient measured in $kg/m^2 . h^\frac{1}{2}$. It is easily determined from the slope of the linear part of the M/A versus $t^\frac{1}{2}$ curve.\\

After a period of sorption the initial rate of ingress observed decreases as the water has accessed all the larger capillary pores. The decrease in gradient of the straight line portion of the water uptake versus square root of time indicates that sorptivity is now occurring via the finer pores and indicates the increasing importance of small pores with time \cite{Martys}.  Besides, some materials with extremely coarse pore structure experience little capillary suction and may show significant deviation from linearity after prolonged wetting.\\
It is widely known that in dry or partially dry mortars or concretes the predominant mechanism in the water absorption is the capillary suction (through the capillary pores) and when the time lapses, and the material begins to be saturated, the predominant mechanism is the diffusion (through the finer and gel pores) \cite{Martys, Xi}. \\
However both processes should coexist simultaneously from early ages (a capillary process through the big capillary pores and a diffusive process through the small pores and gel pores) mainly in slag mortars or concretes where the quantity of gel pores increases and the big capillary pores decrease considerably.\\
To model the capillary and diffusive processes we consider that the cumulative water absorption with time in general can be expressed as:\\

\begin{equation}
\frac{M}{A}=S t^\frac{1}{2} + \mbox{diffusive term}\
\end{equation}

the first term on the right can be written as \cite{Martys}:\\

\begin{equation}
N \rho \lbrace 1-exp( \frac{-S t^\frac{1}{2}}{N \rho})\rbrace\
\end{equation}

where N is a constant related to the distance from the concrete surface over which capillary pores control the initial sorption and $\rho$ is density of the water. This term was constructed so that at early ages, in the limit $t^\frac{1}{2}< \frac {N \rho}{S}$, expansion of the exponential gives $N \rho \lbrace 1-exp( \frac{-S t^\frac{1}{2}}{N \rho})\rbrace\approx S t^\frac{1}{2}$ which is the same as Eq. (1).\\
The diffusive term in (2) is obtained considering the physical diffusion of water and solving the continuity equation or second Fick's law, in non-stationary states with the initial and boundary conditions imposed by the given situation. The samples are cylindrical mortars, permeable only for a face with zero water (considering the water used in the test) as initial condition. In our case all the faces were sealed except the surface in contact with the water. Considering this geometry (Fig. 1) the diffusion problem is reduced to a unidirectional treatment where the diffusion coefficient D is assumed constant for each test.\\

The mathematical problem is given by the equation:\\

\begin{equation}
\frac{\partial C}{\partial t}=D\frac{\partial ^2 C}{\partial x^2}
\end{equation}\\

where C is the water concentration and t the time.\\
The initial and boundary conditions are given by:  \\

\begin {eqnarray}
C(x,0) &=& 0   \:\: para \: x   \in ( 0, L )\\
C(0,t) &=& Co   \:\: para \:     t > 0\\
\frac{\partial C}{\partial x}&=&0\:\: para \:x=L
\end {eqnarray}\\

The first boundary condition (Eq. 6) shows the invariance of water concentration (Co) in the surface for each sample.\\
The second boundary condition (Eq. 7) expresses the impermeability of the top of the sample in x=L.\\
In order to solve the boundary problem the Fourier method \cite{Jost, Carslaw} or the method of laplace transform \cite{Crank} can be used. The result is:\\

\begin{equation}
C(x,t)=Co \{1-\frac{4}{\pi}\:\sum_{n=0}^ \infty\frac{1}{2n+1}\:exp(-\frac{(2n+1)^2\:\pi^2\:Dt}{4 L^2}\:\:\sin\frac{(2n+1)\pi x}{2 L}\}
\end{equation}

This solution expresses the water concentration profiles according to the (x) coordinate and the (t) time in all the section of the sample.\\
The amount of water absorbed through the area A of the permeable surface is:\\

\begin{center}
\begin{eqnarray}
M&=&\int\int\int_V C(x,t)\:dxdydz\\
&=&ACoL\{1-\frac{8}{\pi^2}\:\sum_{n=0}^ \infty\frac{1}{(2n+1)^2}\:exp(-\frac{(2n+1)^2\:\pi^2\:Dt}{4 L^2}\}
\end{eqnarray}
\end{center}

Therefore the amount per unit of area expressed in $kg/m^2$ is given by:\\

\begin{center}
\begin{eqnarray}
\frac{M}{A}=CoL\{1-\frac{8}{\pi^2}\:\sum_{n=0}^ \infty\frac{1}{(2n+1)^2}\:exp(-\frac{(2n+1)^2\:\pi^2\:Dt}{4 L^2}\}
\end{eqnarray}
\end{center}

substituting the Eq. (11) as second term in Eq.(2), we obtain:\\

\begin{center}
\begin{eqnarray}
\frac{M}{A}=N \rho \lbrace 1-exp( \frac{-S t^\frac{1}{2}}{N \rho})\rbrace+CoL\{1-\frac{8}{\pi^2}\:\sum_{n=0}^ \infty\frac{1}{(2n+1)^2}\:exp(-\frac{(2n+1)^2\:\pi^2\:Dt}{4 L^2}\}
\end{eqnarray}
\end{center}

Eq. (12) represents a capillary-diffusive model. This expresses the mass of water per unit of area incorporated to the sample considering both, the capillary sorption through the large pores (first term) and water diffusion through the small pores and very finer gel pores (second term).  \\
\section{Results and Discussion}
  Fig. 2 shows the experimental results of the sorption test for the samples S80W100, S70W100, S60W100, S80W50 and S80A100. A quick decrease of the amount of water absorbed at early ages for all samples is appreciated, showing the samples S80A100 and S80W50 the biggest and smaller rate of absorption at early ages respectively.  Later on the initial rate of ingress observed decreases (the water has accessed all the large capillary pores) until a stabilization of the curve (saturation) is reached for large times in dependence of the analyzed sample.\\
The samples S80W100, S70W100 and S60W100 with 60, 70 and 80\% of slag show a similar behavior. Big differences from the qualitative point of view are not appreciated.
Qualitatively a bigger rate of initial sorption and bigger amount of incorporated water for long times is appreciated for the sample S80A100, followed by S60W100, S70W100 and S80W100. The sample S80W50 shows the smaller rate of initial sorption and the smaller amount of incorporated water for large times.\\
For samples curing at open air (S80A100) a greater rate of uptake of water is qualitatively appreciated in comparison with the samples (S80W100) curing under water. Also, the sample S80W50 preheating to $50^0$C shows a smaller sorptivity and amount of incorporated water that the sample S80W100 preheating to $100^0$C .\\
Shown in figures 3a, 3b, 3c, 4 and 5 are the cumulative water absorption (M/A) versus time (t) for S80W100, S70W100, S60W100, S80W50 and S80A100 samples. Solid lines represent the curves of the fitted model. The fitting of the model (Eq. (12)) permitted us to determine the parameters C, S, D and Co in each case. The values of these parameters for all samples are given in Table 3. In the figures the correlation (r) and multiple determination coefficient ($R^2$) are shown.\\

\begin{center}
\begin{tabular}{|c|c|c|c|c|}\hline\hline
 Mixture & Sorptivity coefficient S  & Diffusion coefficient D & Co& N\\
   &($kg/m^2 s^\frac{1}{2}$) & ($m^2/s$) &($kg/m^3$)& (m)    \\   \hline\hline
S80W100 &$3,16. 10^{-2} \pm 0,25. 10^{-2}$ & $1,38. 10^{-8} \pm 0,08. 10^{-8}$ &$140,52\pm 5,42$
&$1,51.10^{-3}\pm 0,14.10^{-3}$ \\ \hline
S70W100 &$2,75. 10^{-2} \pm 0,24. 10^{-2}$ & $2,50. 10^{-8} \pm 0,23. 10^{-8}$ &$137,34\pm 8,32$
&$1,52.10^{-3}\pm 0,15.10^{-3}$ \\ \hline
S60W100 &$2,81. 10^{-2} \pm 0,15. 10^{-2}$ & $6,10. 10^{-8} \pm 0,27. 10^{-8}$ &$94,62\pm 3,86$
&$4,02.10^{-3}\pm 0,20.10^{-3}$ \\ \hline
S80W50 &$4,80. 10^{-3} \pm 0,45. 10^{-3}$ & $2,83. 10^{-8} \pm 0,14. 10^{-8}$ &$63,75\pm 3,70$
&$2,17.10^{-3}\pm 0,17.10^{-3}$ \\ \hline
S80A100 &$21,31. 10^{-2} \pm 2,1. 10^{-2}$ & $1,28. 10^{-7} \pm 0,08. 10^{-7}$ &$121,52\pm 6,29$
&$2,75.10^{-3}\pm 0,26.10^{-3}$ \\ \hline

\end{tabular}
\normalsize
\end{center}

{\bf Tabla 3}. Values of the parameters for all samples. \\

According with the values of the coefficients S and D the samples S80W100, S70W100, S60W100 with 60, 70 and 80\% slag cured under water and preheating to $100^0$C show a similar behavior which agree with the qualitatively analysis carried out previously. The sample S80W100 curing in water shows a less sorptivity and diffusion (less S and D) in comparison with S80A100 curing at open air and therefore a bigger opposition to the uptake and water transport in the mortars. It is due to the influence of the curing regime on the pore structure. The curing in water facilitates the refinement of the pore structure in slag mortars and therefore leads to high degree of impermeability \cite{Kalid}. In other words, the lack of curing in water in slag cement mortars affects adversely the development of a very finer pore structure. On the other hand the curing at open air increase the absorption and porosity of slag cements \cite{Kelham, Dhir}.\\
The changes in the diffusion coefficients (D) show the role of the diffusive processes, the importance of its consideration as well as their dependence with the structure of pores affected by the different curing regimes. \\
The results obtained for the samples S80W100 and S80W50 allows us to analyze the effect of the pre-conditioning temperature. A diminishing of the sorptivity S (see Table 3) is appreciated when diminish the pre-conditioning temperature. It is due fundamentally to strong role that plays the degree of saturation on the control of water uptake \cite{Martys}.\\
The water flows through the big capillary pores by action of the capillary force, being the flow rate determined by the pore structure and the local content of water. Therefore the capillary forces are bigger in dry materials as S80W100 (preheating at $100^0$C) that in partially dry materials as S80W50 (preheating at $50^0$C) and become minimum in completely saturated materials.\\
It should also be taken into account that for pre-conditioning temperatures around $100^0$C microcracking and other undesirable effects can be generated \cite{Talero, Kelham}, which contributes to a bigger capillary suction. \\
The variation of the diffusion coefficients D with the preconditioning temperature is not significant; being noticed a little increase in the partially saturated sample S80W50 in comparison with the completely dry sample S80W100. \\

\section{Conclusions}

1.  The capillary-diffusive model proposed here describes the kinetics of the water absorption in GGBS mortars with enough accuracy. The kinetic coefficients such as sorptivity and diffusion coefficient should be previously determined in the fitting model process. The model describes the water absorption process for all times and allows us to know the contributions of the diffusion and capillary action in the water transport in mortars.\\
2.  The variation of the \% slag in the range 60-80\% in materials with the same curing regime and preconditioning temperature doesn't influence considerably on the behavior of the materials. This is manifested in the little variation of the coefficients S and D.\\
3.  The curing regime has important effects on the permeability of GGBS mortars. The mortars cured in water show best characteristics of impermeability (less S and D) that the mortars cured at open air.\\
4.  The preconditioning temperature (for the same curing regime) is a factor that shows the influence of the saturation degree on the control of the water uptake. According with the S values, the samples preconditioning at $50^0$C show a less sorptivity in comparison with the samples preconditioning at $100^0$C.\\
5.  The model proposed consider that the capillary sorption and diffusion are the principal mechanisms to the absorption of water in slag cement mortars but this don't consider for simplicity and for having a less contribution other mechanisms (such as permeation, osmosis, thermal migration, etc) that may be present in the uptake and movement of water in these materials. Future studies include the analysis of the possibility of existence (or not) of correlation effects between both, capillary and diffusive process.

\newpage

\begin{center}
{\bf Referencias}
\end{center}

\newpage

\begin{Large}
Figure Captions
\end{Large}

{\bf Figure 1}. Scheme of the sorption test.\\\\\\

{\bf Figure 2}. Cumulative water absorption for all cement mortars added with 60, 70 and 80\% slag\\\\\\

{\bf Figure 3a}. Cumulative water absorption for the sample S80W100 with 80\% slag, curing in water and preconditioned at $100^0$C.\\

               $\bullet$  Experimental $\;\;--$ Model\\\\\\

{\bf Figure 3b}. Cumulative water absorption for the sample S70W100 with 70\% slag, curing in water and preconditioned at $100^0$C.\\

               $\bullet$  Experimental $\;\;--$ Model\\\\\\

{\bf Figure 3c}. Cumulative water absorption for the sample S60W100 with 60\% slag, curing in water and preconditioned at $100^0$C.\\

               $\bullet$  Experimental $\;\;--$ Model\\\\\\

{\bf Figure 4}. Cumulative water absorption for the sample S80A100 with 80\% slag, curing at open air and preconditioned at $100^0$C.\\

               $\bullet$  Experimental $\;\;--$ Model\\\\\\

{\bf Figure 5}. Cumulative water absorption for the sample S80W50 with 80\% slag, curing in water and preconditioned at $50^0$C.\\

               $\bullet$  Experimental $\;\;--$ Model\\\\\\


\begin{enumerate}


\bibitem{Addleson}L. Addleson,  Materials for Buildings, vol. 2 (London: Iliffe Books) p 168 (1972).

\bibitem{Christensen}P. Thoft-Christensen, Deterioration of Concrete Structures, First International Conference on Bridge Maintenance, Safety and Management, LABMAS 2002.

\bibitem{Jensen}O. Mejlhede Jensen, P. Freiesleben Hansen, A. M. Coats, F.P. Glasser, Chloride Ingress in Cement Paste and Mortar, Cement and Concrete Research, Vol.{\bf 29}, N.9, pp.1497-1504 (1999).

\bibitem{Thoft}P. Thoft-Christensen, "What Happens with Reinforced Concrete Structures when the Reinforcement Corrodes", Proc. I. Workshop on "Life-Cycle Cost Analysis and Design of Civil Infrastructure Systems", Ube, Yamaguchi, Japan, September 2001.

\bibitem{Luping}T.Luping, "Chloride Transport in Concrete - measurements and prediction", Diss. Chalmers University of Technology, Department of Building Materials, Publication P-96: 6, (1996).

\bibitem{Mironova}M. Mironova, P. Gospodinov, R. Kazandjiev, The effect of liquid push out of the materials capillaries under sulfate ion diffusion in cement composite, Cem.Concr.Res {\bf 32} (2002) 9-15.

\bibitem{Martys}N. Martys, C. Ferraris, Capillary transport in mortars and concretes, Cem.Concr.Res {\bf 27}, (1997) 747-760.

\bibitem{1Hall}C. Hall, T. Kam Min Tse, Water movement in porous building materials - VII: The sorptivity of mortars, Bldg. Env. {\bf 21} (1986) 113-118.

\bibitem{Ho}D.W.  Ho, R.K. Lewis, The water sorptivity of concrete: the influence of constituents under conditions curing, Durab.Buildg.Mater., {\bf 4} (1987) 241-252.

\bibitem{2Hall}C. Hall, M.H. Yau, Water movement in porous building materials - IX: The water absorption and sorptivity of concretes, Ibid {\bf 22} (1987) 77-82.

\bibitem{3Hall}C. Hall, Water sorptivity of mortars and concrete: a review, Mag.Concrete.Res. {\bf 41} (1989) 51-61.

\bibitem{Farrell}B. Sabir, S. Wild, M. O`Farrell, A water sorptivity for mortar and concrete, Materials and Structures {\bf 31} (1998) 568-574.

\bibitem{Ohdaira}E. Ohdaira, N. Mazuzawa, Water content and its effect on ultrasound propagation in concrete - the possibility of NDE, Ultrasonics {\bf 38} (2000) 546-552.

\bibitem{Goual}M.S. Goual, F. de Barquín, M.L. Benmalek, A. Bali, M. Quéneudec, Estimation of the capillary transport coefficient of Clayed Aerated Concrete using a gravimetric technique, Cem.Concr.Res. {\bf 30} (2000) 1559-1563.


\bibitem{McCarter1}W. J. McCarter, D. W. Watson, T.M. Chrisp, Surface zone concrete: Drying, Absorption and Moisture Distribution, J. of Mater. in Civ.Engin, January (2001).

\bibitem{McCarter2}W. J. McCarter, T.M. Chrisp, A. Butler, P.A.M. Basheer, Near-surface sensors for condition monitoring of cover-zone concrete, Const.Bldg.Mater. {\bf 15} (2001).

\bibitem{Swamy}R.N. Swamy, Design for durability and strength through the use of fly ash and slag in concrete, Mario Collepardi Symposium on Advances in Concrete Science and Technology, P.K. Meta Ed., Rome 8 October (1997) 127-194.

\bibitem{Metha}P.K. Metha, Concrete Structure, Properties and Materials, Prentice-Hall, Inc., Englewood Cliffs, N.J. 07632, (1993).

\bibitem{Klieger}P. Klieger, Joseph F. Lamond (Eds), Significance of tests and properties of concrete and concrete-making materials, ASTM STP 169C, (1994).

\bibitem{Metha1} P.K. Metha, Pozzolanic and cementitious byproducts as mineral admixtures for concretes- A critical rewiev, ACI Publ. SP-79, ed. V.M. Malhortra vol. 1, (1983), 1-46.

\bibitem{Kalid} M. Kalid, S.P. Mehrotra, M. Verma, J. Ahmad, C. Verma, Performance of multiblend cement under aggressive environment, The Indian Concr. J. January (2002), 22-26.

\bibitem{Mejia}R. Mejia, S. Delvasto, R. Talero, Performance of GGBS cements, 13th International Conference on solid waste technology and management Proceedings, Philadelphia, USA (1997).

\bibitem{Talero}R. Mejia, R. Talero, Effect of the curing and preconditioning regime on the absorptivity of added mortars, 2nd NACE Latin American Corrosion Congress, September 1996.

\bibitem{Thomas}M.D.A. Thomas, P.B. Bamforth, Modelling chloride diffusion in concrete. Effect of fly ash and slag, Cem,Concr.Res., {\bf 29} (1999) 487-495.

\bibitem{Ho1}D.W. Ho, Influence of slag cement on the water sorptivity of concrete, in "Fly ash, silica fume, slag and natural pozzolans in concrete", Proceedings of an International Conference (American Concrete Institute, SP-91, Detroit) (1986) 1463-1473.

\bibitem{Kuntz}M. Küntz, P. Lavallé, Experimental evidence and theoretical analysis of anomalous diffusion during water infiltration in porous building materials, J. Phys.D: Appl.Phys. {\bf 34} (2001) 1-8.

\bibitem{Holm}A. Holm, H. Künzel, Non-isothermal moisture transfer in porous building materials, Materialsweek, September 2000-Munich.

\bibitem{Martys2}N. Martys, Diffusion in partially-saturated porous materials, Materials and Structures, {\bf 32} (1999) 555-562.

\bibitem{Claisse}P. Claisse, H. Elsayad, I. Shaaban, Absorption and sorptivity of cover concrete, J. of Mater. in Civ.Engin, August (1997).


\bibitem{Xi}Y. Xi, Z.P. Bazant, L.Molina, H.M. Jennings, Advn.Cem.Bas.Mat., {\bf 1}, 258, (1994).

\bibitem{Jost}W. Jost, Diffusion in solids, liquids, and gases, Academic Press, New York, (1960).

\bibitem{Carslaw}H.S. Carslaw, and J. C. Jaeger, Conduction of Heat in Solids, Oxford University Press, Oxford (1970).

\bibitem{Crank}J. Crank, The Mathematics of Diffusion, Clarendon Press, Oxford (1975).

\bibitem{Kelham}S. Kelham, Magazine of Concrete Research, {\bf 40} (1988) 106-110.

\bibitem{Dhir}R.K. Dhir, E.A. Byars, Magazine of Concrete Research, {\bf 43} (1991) 219-232.


\end{enumerate}
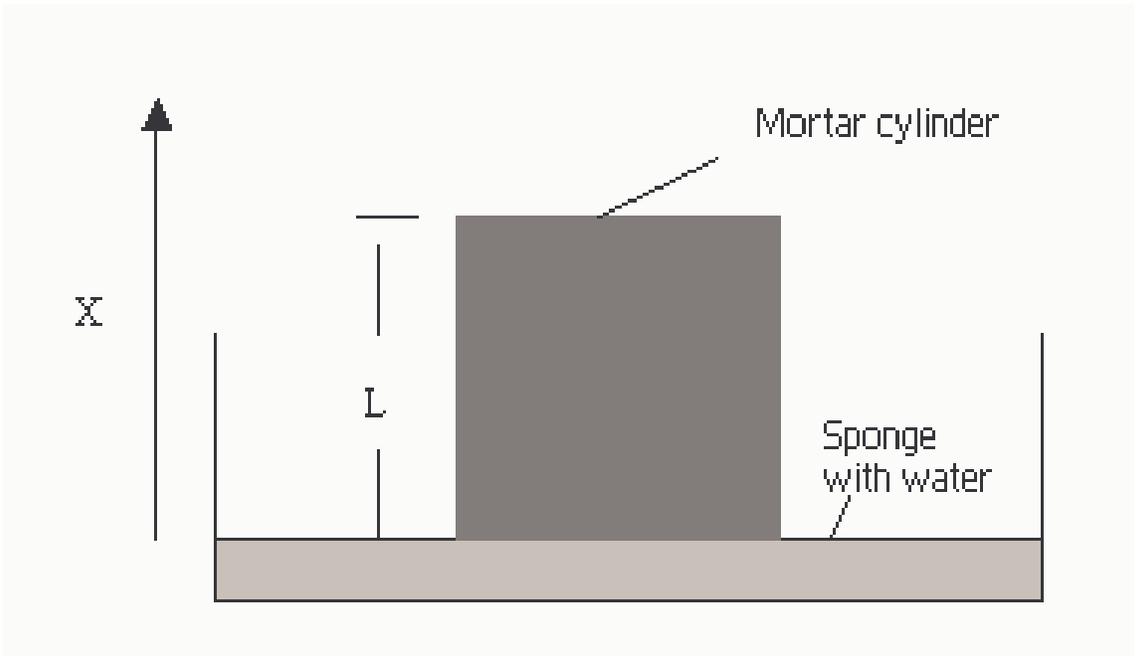
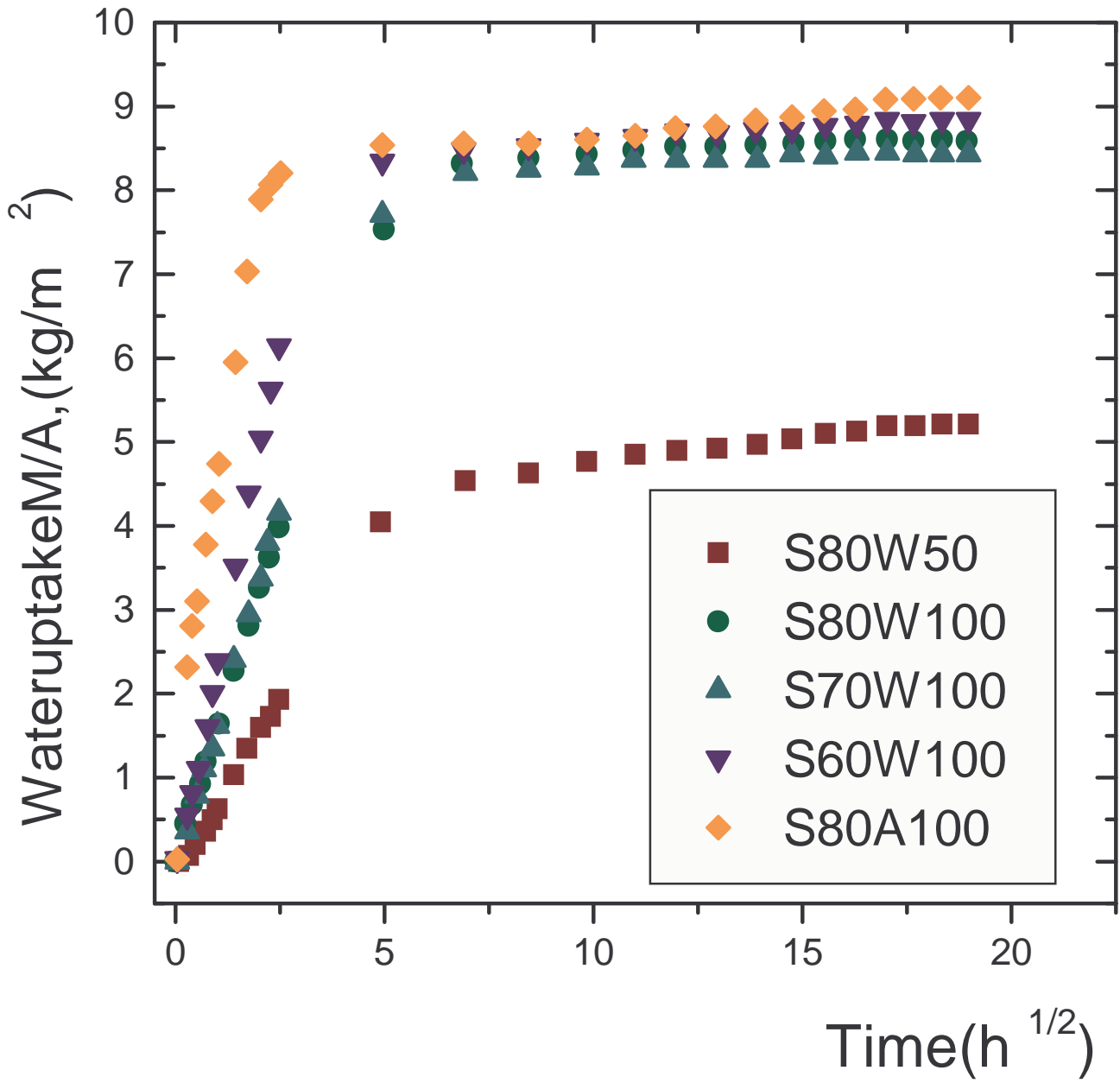
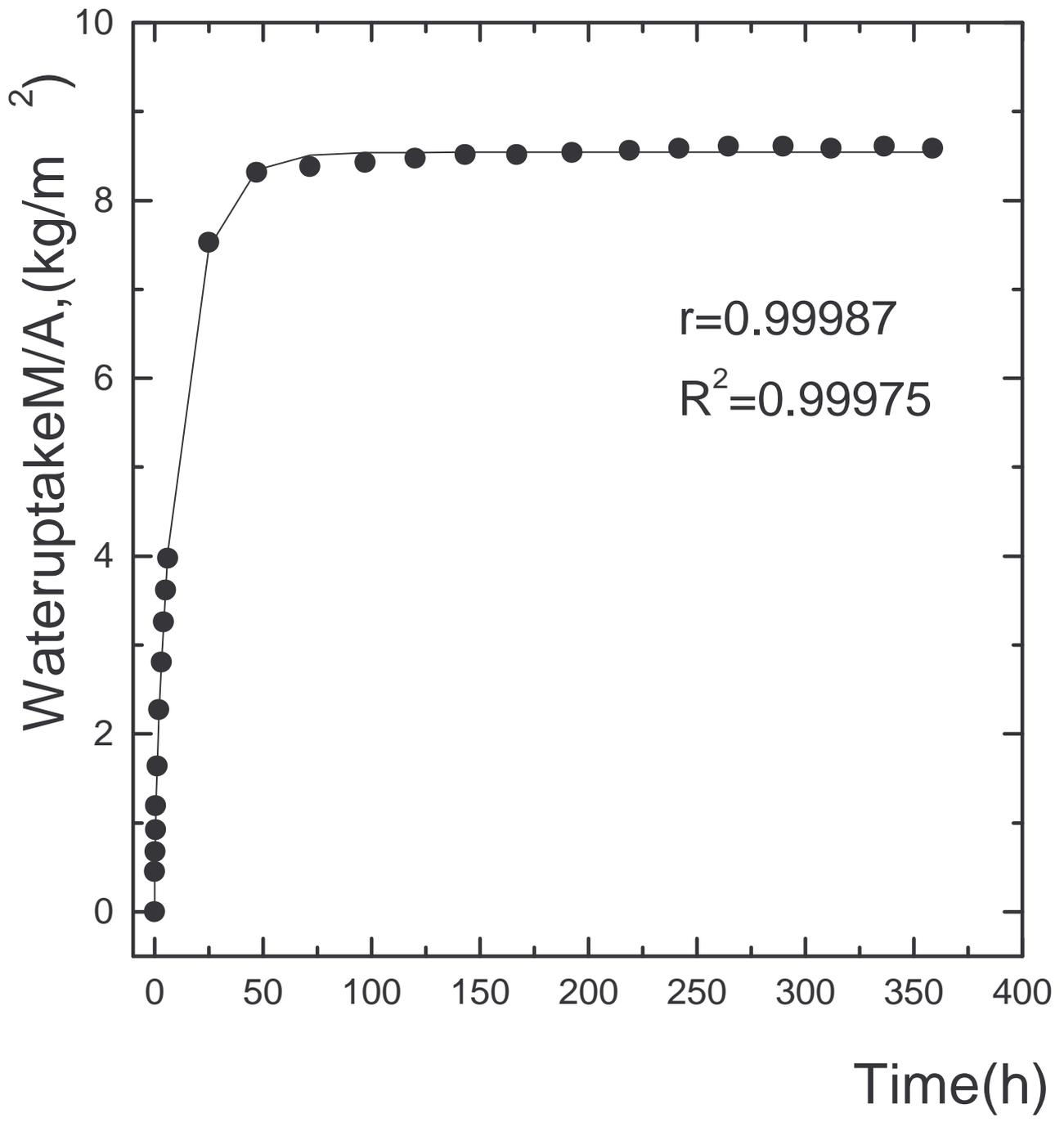
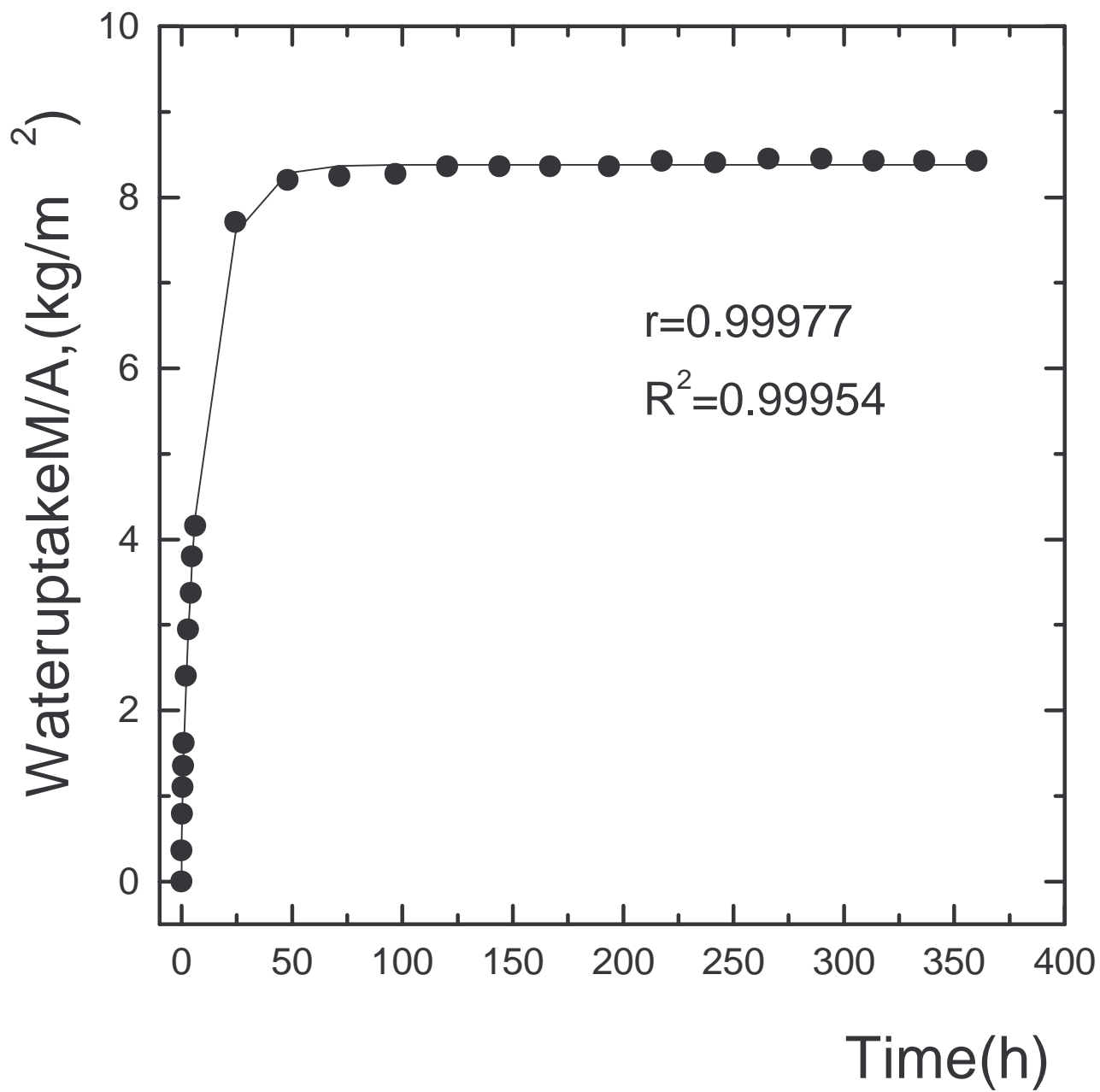
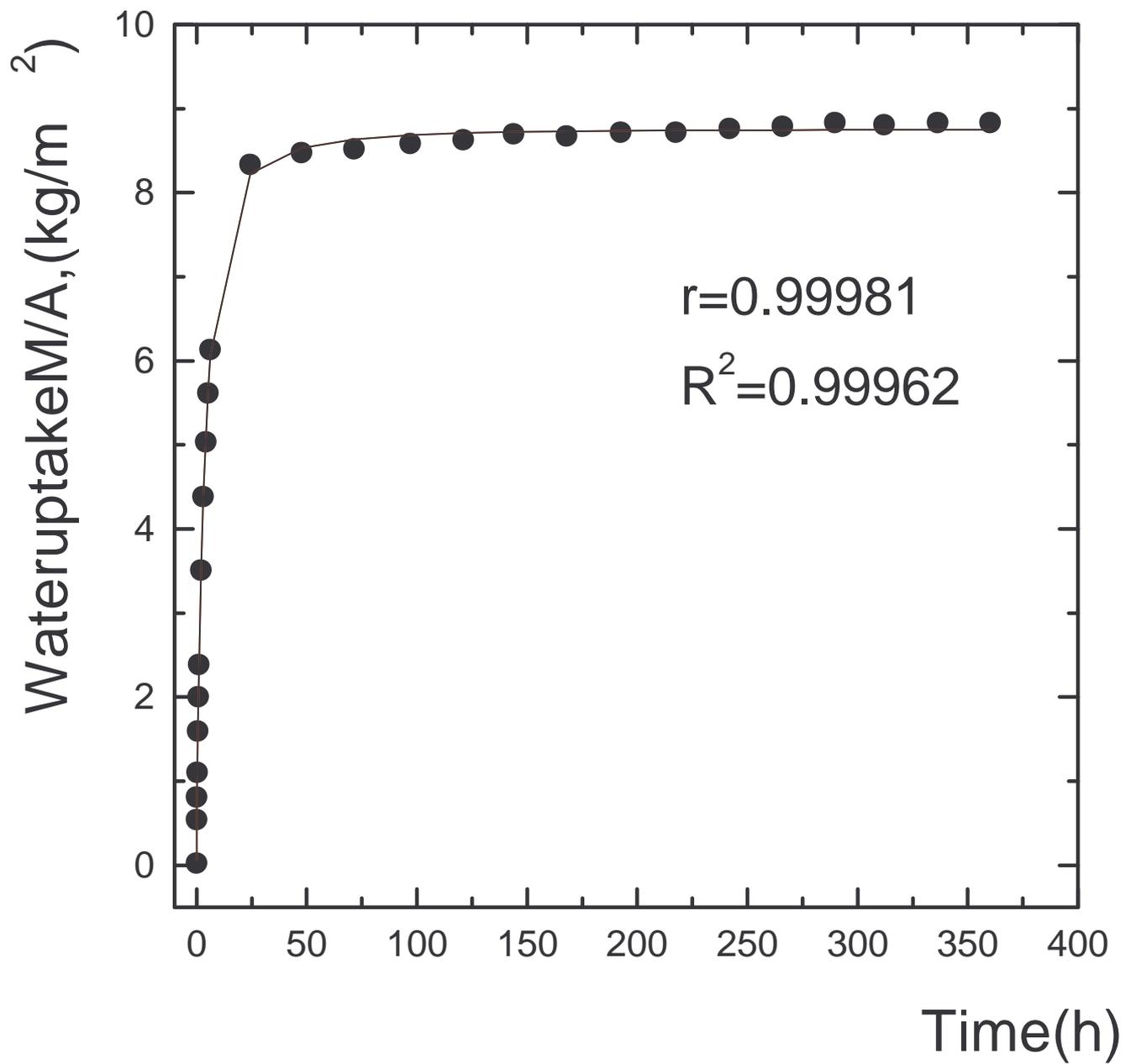
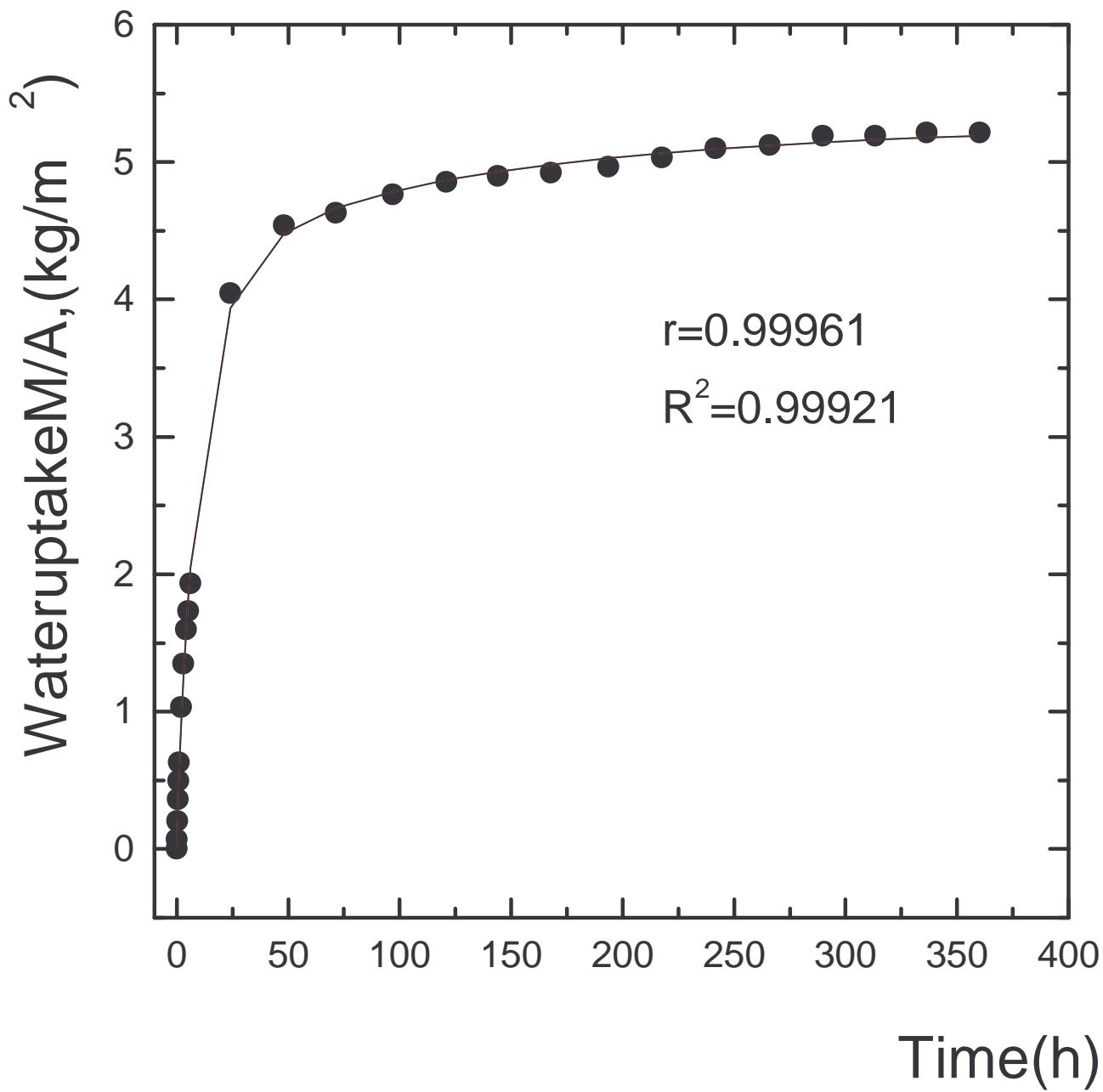
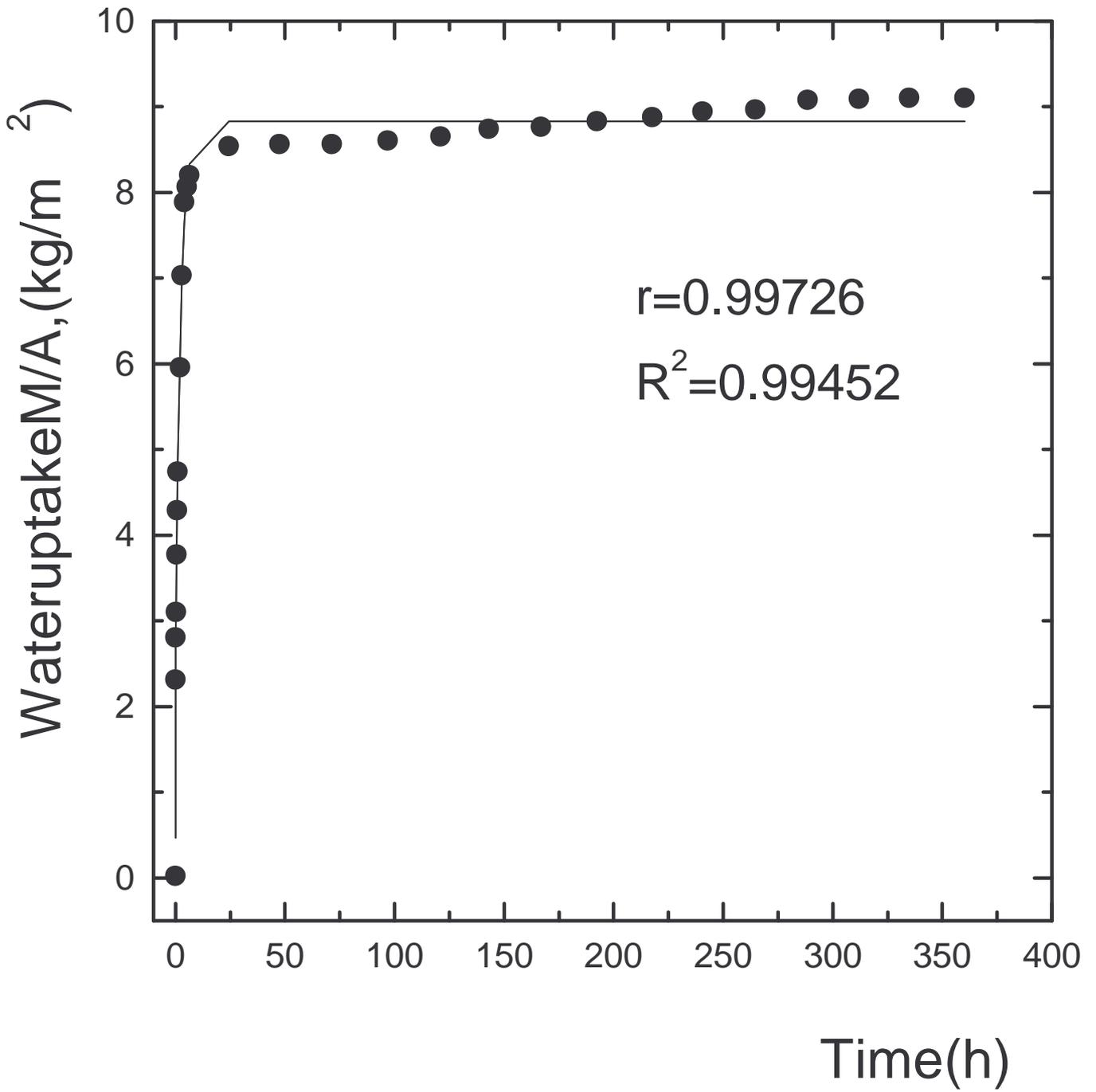
\end{document}